\documentclass{article}

\usepackage{arxiv}

\usepackage{latexsym}
\usepackage{graphicx}
\usepackage{multicol,multirow}
\usepackage{amsmath,amssymb,amsfonts}
\usepackage{mathrsfs}
\usepackage{amsthm}
\usepackage{algorithmic}
\usepackage{algorithm}
\usepackage{apacite}
\usepackage{rotating}
\usepackage{appendix}
\usepackage[authoryear]{natbib}
\usepackage{ifpdf}
\usepackage[T1]{fontenc}
\usepackage{times}
\usepackage{sourcesanspro}
\usepackage{newtxmath}
\usepackage{textcomp}%
\usepackage{xcolor}%
\usepackage{hyperref}
\usepackage{caption}
\usepackage{subcaption}

\usepackage[utf8]{inputenc} 
\usepackage[T1]{fontenc}    
\usepackage{url}            
\usepackage{booktabs}       
\usepackage{nicefrac}       
\usepackage{microtype}      
\usepackage{lipsum}		

\defcitealias{AIS-IMO:2003}{IMO, 2003}

\title{An optical flow approach to tracking ship track behavior using GOES-R satellite imagery}

\author{ \href{https://orcid.org/0000-0002-1731-3834}{\includegraphics[scale=0.06]{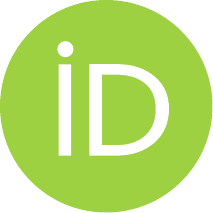}\hspace{1mm}Kelsie M. Larson}
\thanks{corresponding author, kmlarso@sandia.gov} \\
	Sandia National Laboratories\\
	Albuquerque, NM \\
	\And
	\href{https://orcid.org/0000-0002-5239-5185}{ \includegraphics[scale=0.06]{orcid.pdf}\hspace{1mm}Lyndsay Shand} \\
	Sandia National Laboratories\\
	Albuquerque, NM \\	\\
	University of Illinois, Urbana-Champaign\\
	Urbana, Illinois\\
	\And
	\href{https://orcid.org/0000-0001-6125-5814}{ \includegraphics[scale=0.06]{orcid.pdf}\hspace{1mm}Andrea Staid} \\
	Electric Power Research Institute\\
	Palo Alto, CA \\	
	\And
	Skyler Gray\\
	Sandia National Laboratories\\
	Albuquerque, NM \\	\\
	Brigham Young University\\
	Provo, UT\\
	\And
	\href{https://orcid.org/0000-0001-9702-1737}{\includegraphics[scale=0.06]{orcid.pdf}\hspace{1mm}Erika L. Roesler}\\
	Sandia National Laboratories\\
	Albuquerque, NM \\	
	\And
	Don Lyons\\
	Sandia National Laboratories\\
	Albuquerque, NM \\	
}

\renewcommand{\shorttitle}{\textit{arXiv} Template}

\begin{document}
\maketitle

\begin{abstract}
Ship emissions can form linear cloud structures, or \textit{ship tracks}, when atmospheric water vapor condenses on aerosols in the ship exhaust.  
These structures are of interest because they are observable and traceable examples of marine cloud brightening, a mechanism that has been studied as a potential approach for solar climate intervention. 
Ship tracks can be observed throughout the diurnal cycle via space-borne assets like the Advanced Baseline Imagers on the National Oceanic and Atmospheric Administration Geostationary Operational Environmental Satellites, the GOES-R series. Due to complex atmospheric dynamics, it can be difficult to track these aerosol perturbations over space and time to precisely characterize how long a single emission source can significantly contribute to indirect radiative forcing. 
We propose an optical flow approach to estimate the trajectories of ship-emitted aerosols after they begin mixing with low boundary layer clouds using GOES-17 satellite imagery. Most optical flow estimation methods have only been used to estimate large scale atmospheric motion. We demonstrate the ability of our approach to precisely isolate the movement of ship tracks in low-lying clouds from the movement of large swaths of high clouds that often dominate the scene. This efficient approach shows that ship tracks persist as visible, linear features beyond 9 hours and sometimes longer than 24 hours.
\end{abstract}

\keywords{ship tracks \and optical flow \and HYSPLIT \and cloud-aerosol interactions, GOES-R}

\section{Introduction}
It is well documented that aerosols from anthropogenic sources can apply direct radiative forcing by reflecting or absorbing sunlight, as well as apply indirect radiative forcing by altering the radiative properties of low-lying clouds \cite{twomey1974, albrecht1989, Seinfeld2016, Christensen2020}. 
The magnitude of the impact of aerosols on a cloud's radiative properties can vary greatly depending on the properties of the aerosol and the surrounding atmosphere \cite{chen2014}. Most often, anthropogenic aerosols increase the amount of radiation reflected by clouds, but in some cases, they have been known to reduce a cloud's albedo \cite{chen2012}. 

Currently, indirect radiative forcing is the largest documented source of uncertainty when it comes to overall radiative forcing in climate modeling \cite{carslaw2013, icc2013, wang2020}. 
This large uncertainty is due in part to the complexity of cloud dynamics, making it difficult to clearly separate the aerosol's radiative effect from that of the surrounding clouds \cite{stevens2009}. Improving our understanding of aerosol-cloud interactions is necessary to reduce this uncertainty in climate models. 
An important implication of these interactions, increasing the reflectivity of clouds, has the potential to reduce positive radiative heating through targeted ``climate cooling,'' the central concept in solar climate intervention.
Marine cloud brightening (MCB) and other intervention approaches have been proposed to intentionally increase the reflectivity of low altitude, boundary-layer clouds, e.g. \cite{latham1990, ci2015, icc2013}.

Ship emissions have provided researchers with observable and traceable examples of aerosol-cloud interactions, which have been the focus of many studies to better understand the potential impacts of MCB \cite{hobbs2000, glassmeier2021}.
For more than fifty years, satellite imagery has detected these bright linear features produced when the engine exhaust from large ocean-traversing ships mixes with low-lying marine clouds within 2 km of the earth's surface \cite{conover1966, twomey1968}.
Ship tracks were first observed in visible-wavelength images taken from the Television Infrared Observational Satellites (TIROSs) and the Monterey Area Ship Track experiment off the coast of California \cite{durkee2000} was one of the largest aircraft campaigns to study the formation of ship tracks.

A lack of high resolution data, as well as difficulties isolating and tracing observed aerosol-cloud interactions over time, have been limiting factors in studying the longevity and long-term effects of ship tracks. 
Costly air campaigns have been the most reliable method of tracking the behavior of aerosols from a known source.
The improved spatial, spectral, temporal, and radiometric resolutions of new generation imagers enable new methods for aerosol tracking.
For example, \cite{zhao2018} used the Advanced Very High Resolution Radiometer (AVHRR) from the National Oceanic and Atmospheric Administration (NOAA) satellites to study the global long-term indirect effects of aerosols.
\cite{yuan2019} used machine learning to automatically label ship tracks in images from the MODerate resolution Imaging Spectroradiometer (MODIS) aboard both the Aqua and Terra satellites.
\cite{gryspeerdt2019} combined MODIS imagery and retrievals of cloud droplet number concentration with known ship positions and properties to demonstrate a positive effect of emission sulfate concentration on the likelihood of ship track formation and a decrease in ship track observations due to fuel sulfur content restrictions set by the International Maritime Organization (IMO).
More recently, \cite{diamond2020} applied spatial kriging methods to cloud property data retrieved from MODIS and reanalysis from the Modern‐Era Retrospective analysis for Research and Applications, Version 2 (MERRA‐2), to infer negative impacts on radiative forcing from emissions along a major shipping lane in the southeast Atlantic.
New satellite observations of aerosol-cloud interactions have been a large source of untapped information since it is very difficult to infer the radiative impact of ship emissions from observational data collected from earlier generations of satellites \cite{falkowski1992, kim1993, han1994}.

Historically, our understanding of the interactions between injected aerosols and clouds has been primarily limited to simulations.
In atmospheric computational fluid dynamic numerical models, aerosol injections are initiated in the model at known, precise locations in fully defined environments that are easily traceable, e.g. \cite{wang2011, berner2015, possner2018, blossey2018}.
Unlike simulated case studies, satellite observations of ship tracks have many uncertainties. 
Real, observed ship tracks are initiated from an often-unknown source and form in a dynamic and only partially known environment, making it challenging to trace and fully characterize their behavior. 
For a known air parcel, NOAA's Hybrid Single-Particle Lagrangian Integrated Trajectory (HYSPLIT) model represents the state of the art in tracking
atmospheric movement \cite{hysplit}. This atmospheric transport and dispersion model is widely used by atmospheric scientists to estimate and study the trajectories of air parcels forward and backward in time. This can be applied to a ship track to attempt to follow its movement and evolution, thus determining the duration of radiative impact. While 
HYSPLIT captures broad trajectory movement well at coarse hourly intervals, the errors are often too large to be able to follow a small, unique structure, such as a ship track, with confidence.  

In the research we present herein, we show how high-resolution spatial and temporal observations from the NOAA GOES-17 Advanced Baseline Imager (ABI)\cite{goesrabi} data can resolve ship tracks day and night. 
We propose and demonstrate the use of an optical flow algorithm, which relies on pixel information alone, to effectively follow the behavior and persistence of ship tracks in space and time. By doing so, we show that this method can be used to determine how long ship tracks persist in a maritime 
environment -- a key question at the heart of understanding broader aerosol impacts for MCB. This proposed method could be used as a means to record ship track impacts on climate. 

``Optical flow is the distribution of apparent velocities of movement of brightness patterns in an image. Optical flow can arise from relative motion of objects and the viewer" \cite{horn1981}. For example, the GOES Atmospheric Motion Vectors are constructed using Patch Matching (PM) to estimate the optical flow from observed cloud- and water-vapor drift motion \cite{fortun2015}. While PM is computationally efficient and effective at tracking coherent textures with large displacements, it breaks down when tracking scenes with little texture, strong deformations, discontinuities, transparent motions, and illumination changes. Hence, GOES Atmospheric Motion Vectors implement strict quality control pruning procedures and return sparse, synoptic-scale winds that are primarily designed to be ingested by numerical models.
With recent improvement in optical sensors on satellite systems, dense optical flow methods have been proposed to estimate smaller-scale displacements when the data provides sufficiently high temporal resolution such as the 5--15 minute intervals produced by the GOES-17 ABI, e.g. \cite{horn1981, sun2014, apke2020}. However, these methods retrieve optical flow by blending local information with regional flow estimates through regularization that often smooths away motions from small and isolated features, particularly those with large displacements \cite{malik2011, fortun2015}.
Our optical flow estimation method similarly leverages the high temporal resolution of the GOES-R ABI and relies on the image registration technique of \cite{lk1981}, which computes optical flow for a sparse feature set, within a subsetted image window. Whereas dense optical flow methods require pixel matching of all pixels within a scene, sparse optical flow methods only require the processing of a select set of pixels representing the feature(s) of interest. This is more ideal for tracking small, local regions, such as in the case of ship tracks where we are interested in isolating the motion of local low-lying clouds from the motion of surrounding high clouds.

Our method uses radiance spectra collected from the GOES-17 ABI sensor \cite{goesrabi} and provides an efficient approach to accurately and systematically follow a specific track and characterize ship track persistence. We use the HYSPLIT model as a baseline comparison and present performance results in terms of tracking accuracy for known ship track segments. 
We have observed a large variation in terms of track persistence, and we show examples of tracks that persist as detectable linear cloud features for as long as 12 to 24 hours, much longer than the 6--8 hr typically assumed in climate simulation studies in pristine environments, e.g. \cite{berner2015}. The remainder of this paper is organized as follows: Section \ref{sec:data} outlines our data sources; Section \ref{sec:imageflow} describes our optical flow approach; Section \ref{sec:hysplit} compares our approach to HYSPLIT,  a standard tool for the trajectory analysis of aerosols; and Section \ref{sec:results} assesses the performance of both methods. Finally, Section \ref{sec:disc} discusses how our proposed approach might be used in practice, and recommends follow-on work.

\section{Data}
\label{sec:data}

\noindent This research uses L1b radiances measured from the Advanced Baseline Imager (ABI) instrument on the GOES-17 geostationary weather satellite \cite{goesrabi}, which provides four times higher spatial and three times higher temporal resolutions than previous generations of GOES imagers. Higher resolutions in both space and time allow us to study fast-changing cloud structures such as ship tracks with greater precision. We rely on the near-infrared ``cloud particle size" band (C06) and the infrared ``shortwave window" band (C07) with central wavelengths of 2.24 and 3.90 $\mu m$, respectively, to visualize ship tracks throughout the diurnal cycle. To seamlessly visualize ship tracks during day-night transitions, we transform the data for each time stamp by subtracting spectral band C07 from band C06 and apply the image processing technique of histogram equalization to systematically control the contrast of each image. Histogram equalization enhances the contrast of each image, making it easier to visualize and identify key ship track features that may otherwise be invisible to the naked eye. The spatial resolution of both bands (C06 and C07) is 2 km, and the temporal resolution is every five minutes for the GOES-17 CONUS scan.


For the study presented here, we examined GOES-17 CONUS satellite imagery for a selection of dates with clearly visible track examples. We then selected 24 examples of track intersections for further case study. These examples consist of ship tracks in the Northern Pacific Ocean, often some distance off the western coast of the U.S. in February, April, and June 2019. We focused the study on track intersections because they are visibly identifiable over time and can provide an initial qualitative assessment of our tracking method.
Each example has a different composition of high and low clouds and thus exhibited different ship track behaviors (i.e., different feature formations and movements). While not a comprehensive study, these examples demonstrate the robustness of our approaches under different ship track conditions. The exact locations and timestamps of the initial formation of these track intersections are detailed in Table \ref{tab:intersections}.

\begin{table}[!t]
	\caption{Details of manually selected ship track intersections. All tracks were identified from the CONUS scan of the GOES-R ABI and thus reside over the North Pacific Ocean. Initialization heights are inferred from the ERA-5, the fifth generation European Centre for Medium-Range Weather Forecasts ECMWF reanalysis. For more details see \url{https://www.ecmwf.int/en/forecasts/datasets/reanalysis-datasets/era5}}
	\label{tab:intersections}
	\centering
	\resizebox{\columnwidth}{!}{
	\begin{tabular}{|c|c|c|c|c|}
		\hline
		\textbf{Date (MM/SS/YY)} & \textbf{Time (HH:MM:SS UTC)} & \textbf{Latitude ($^\circ$ N) } & \textbf{Longitude ($^\circ$ W) } & \textbf{Height (m)}\\ \hline
		
			02/20/19 	& 00:02:38 	& 40.31 	& 145.71  	& 573 \\
			02/20/19 	& 15:02:38 	& 44.83 	& 140.07  	& 945\\
			02/20/19 	& 18:02:38 	& 39.42 	& 138.13  	& 204\\
			02/20/19 	& 18:02:38 	& 43.52 	& 139.29  	& 1190\\
			02/20/19 	& 22:02:38 	& 39.37 	& 140.45  	& 178\\
			04/24/19 	& 01:02:38 	& 35.50 	&128.44  	& 449\\
			04/24/19 	& 04:02:38 	& 34.48 	& 128.54  	& 389\\
			04/24/19 	& 04:02:38 	& 38.03 	&128.00  	& 552\\
			04/24/19 	& 04:02:38 	& 38.61 	& 128.43 	& 642\\
			04/24/19 	& 07:02:38 	& 35.06 	&129.37  	& 393\\
			06/2/19 	& 00:02:38 	& 36.79 	&132.26  	& 549\\
			06/2/19 	& 00:02:38 	& 33.75 	& 128.53 	& 572\\
			06/2/19 	& 12:07:38 	& 34.54 	&129.86  	& 585\\
			06/2/19 	& 12:02:38 	& 33.87 	&128.88  	& 572\\
			06/2/19 	& 17:02:38 	& 34.11 	& 128.92  	& 572\\
			06/2/19 	& 17:02:38 	& 34.24 	& 128.27  	& 573\\
			06/2/19 	& 17:02:38 	& 33.35 	& 127.90  	& 601\\
			06/17/19 	& 00:02:38 	& 36.42 	& 130.58  	& 620\\
			06/17/19 	& 07:02:38 	& 36.19 	&134.25  	& 589\\
			06/17/19 	& 09:02:38 	& 35.51 	& 134.30  	& 589\\
			06/17/19 	& 09:02:38 	& 35.36 	&131.16 	& 588\\
			06/17/19 	& 09:02:38 	& 34.79 	& 131.03 	& 588\\
			06/17/19 	& 11:02:38 	& 35.37 	& 130.58  	& 588\\
			06/17/19 	& 11:02:38 	& 34.56 	&130.31  	& 578\\ \hline
	\end{tabular}
	}
\end{table}


\section{Tracking ship tracks with optical flow}
\label{sec:imageflow}

\noindent In this section we present an optical flow approach to reliably track ship track features across more than 24 hr of consecutive GOES-17 satellite image frames.  Our approach relies on the Shi-Tomasi (ST) algorithm, also known as \texttt{Good Features to Track} \cite{st1994}, to select trackable features and a pyramidal implementation of the Lucas-Kanade (LK) algorithm \cite{jyblk} to estimate the optical flow, or the feature displacement, between image frames.  This pair of algorithms is widely used in the video processing domain to track moving objects; the LK algorithm alone has been previously used for cloud motion estimation, both in ground-based video feeds to forecast solar irradiance \cite{bzp2012} and in satellite image sequences to track individual cloud banks \cite{il2013}.  The LK algorithm has been shown to be most successful with high-contrast textural features and the ST algorithm is designed specifically to select such features \cite{st1994}, making the algorithm pair an appropriate choice for tracking the textured, cloudy regions where ship tracks are observed.  The GOES-R CONUS scan data used are also well-suited for this tracking application as the high temporal resolution (consecutive frames are 5 minutes apart) allows the major assumptions of LK tracking - that feature displacement is relatively small and feature content remains relatively constant - to hold.  Though the shapes of ship tracks evolve over time, the changes are sufficiently minimal over the short frame duration to enable tracking.

Although dense optical flow methods, like those in \cite{horn1981, fortun2015, sun2014, baker2011}, may also apply to this problem, the sparse, pyramidal LK algorithm was selected due to its simplicity, efficiency, and the availability of open source implementations.  We found the LK algorithm to be an appropriate choice for our initial exploration of ship track persistence, especially as it has shown success in a range of applications \cite{bzp2012, il2013, paugam2021, ivanov2022, TLD2012, choi2022, xiang2015, kotsia2007, bojsen2005, kanhere2008, matsushita2005}, and we leave the application of other dense flow methods to future studies.

This approach relies on image pixel values alone with minimal consideration of other atmospheric data and is therefore unaffected by the meteorological errors and uncertainties to which physical trajectory models such as HYSPLIT are subjected.  Thus, it is an attractive method for tracking cloud features observed at any altitude, but like most optical flow algorithms, it is sensitive to image data corruption and intensity variation between frames. Intensity variation is common between GOES-17 CONUS frames, especially when transitioning between nighttime and daytime images. As described in this section, our optical flow method addresses this specific issue to enable continuous feature tracking for more than 24 hr.

Our algorithm implementation is available for open source use\footnote{Our optical flow algorithm is available for open source use at: https://github.com/sandialabs/CFTrack}.  Note that the user is required to download the appropriate GOES L1b radiance imagery and organize it as described in the instructions provided with the code.

\subsection{Optical Flow Method Description}
\begin{figure*}[!t]
	\centering
	\subfloat[Initialization: 0 Hours]{\includegraphics[width=2.75in]{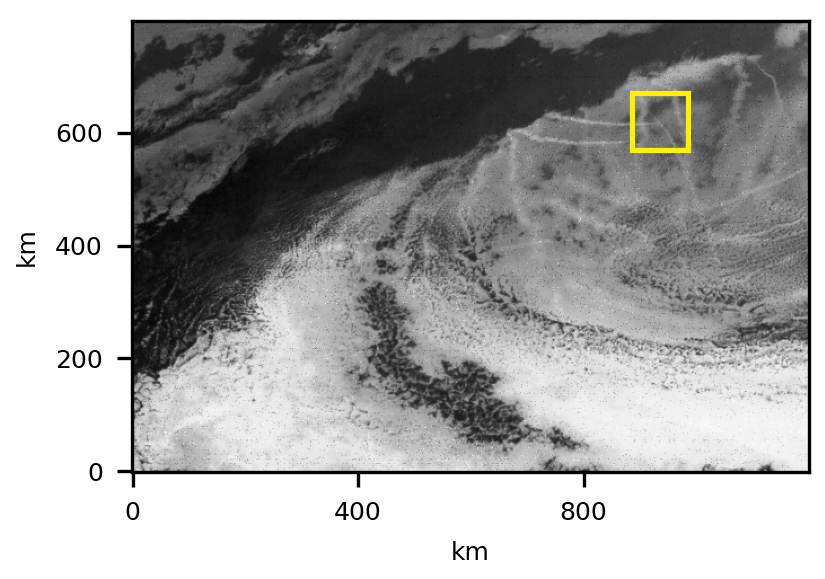}}
	\subfloat[6 Hours]{\includegraphics[width=2.75in]{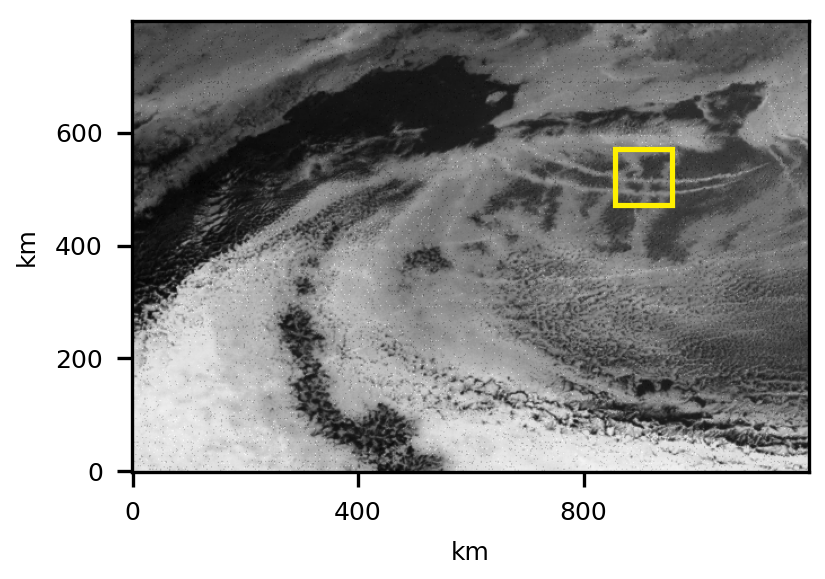}}\\
	\subfloat[12 Hours]{\includegraphics[width=2.75in]{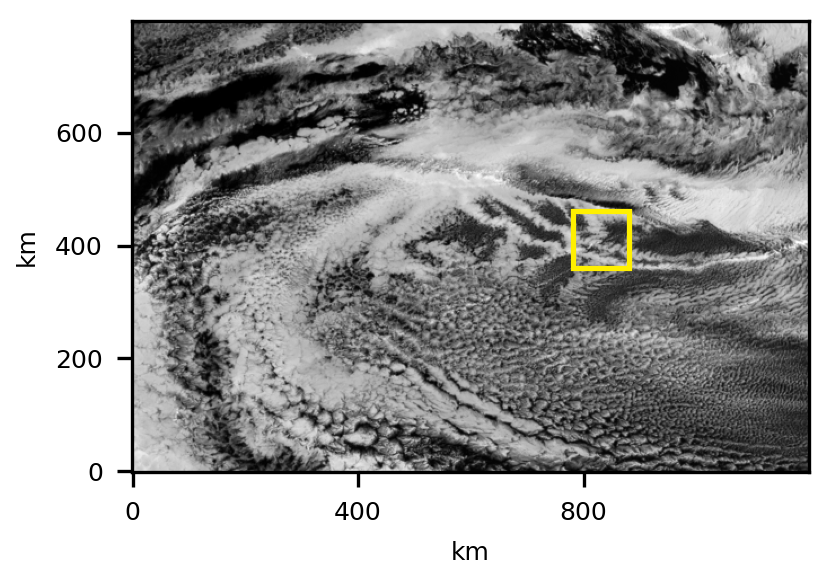}}
	\subfloat[18 Hours]{\includegraphics[width=2.75in]{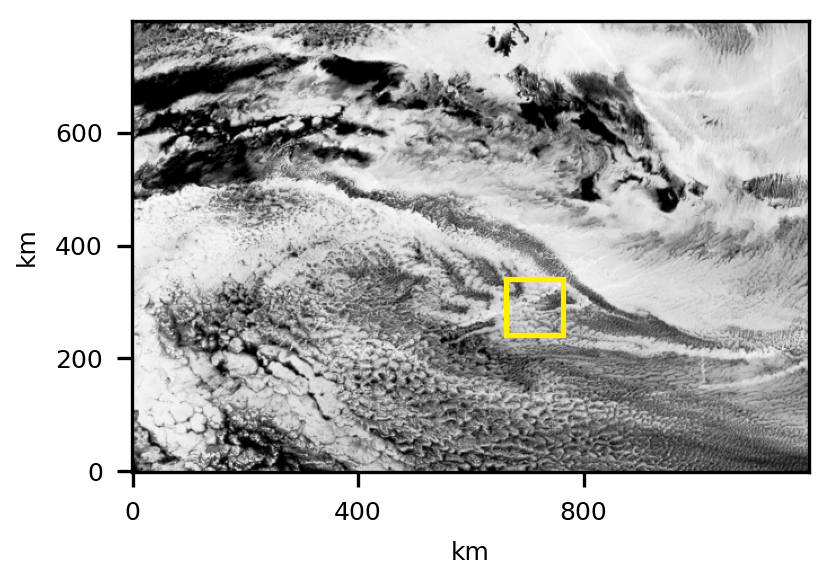}}
	
	\caption{These figures show the result of the optical flow method applied to a manually-selected local cloud region, starting with an intersection of two ship tracks on June 17, 2019, at 07:02 UTC (a) and stepping forward in time, with snapshots shown at 6 (b), 12 (c), and 18 (d) hr later. The tracking algorithm is able to follow the movement of the cloud region well, and the tracks are still clearly visible 18 hr later.  The center location of these images is 33°27'02.0"N 138°06'11.9"W.}
	\label{fig:lk_june_big}
	
\end{figure*}

The ST and LK algorithms are repeatedly applied to track local cloud regions containing ship tracks in the following manner.  First, we manually select an image region of interest (ROI) immediately surrounding an intersection of two ship tracks, and bound it with a ``tracking box'' (see, for example, Fig. \ref{fig:lk_june_big}).  This tracking box allows for the isolation of movement of the lower cloud layer where the ship track forms and ignores high cloud movement which often dominates a scene. We choose to track intersections for demonstration as they remain visually distinctive over time, but this method could also be applied to other textured cloud regions or ship track ``heads,'' positions at which new, visible cloud track is forming.  We choose the size of the ROI and associated tracking box such that at least five high-contrast features, as selected by the ST algorithm in the next step, could be found within the region; we found that an approximately 50 $\times$ 50 pixel or 100 $\times$ 100 km region often satisfied this constraint.

We next apply the ST algorithm within the ROI in the first frame to identify trackable features.  We consecutively apply the LK algorithm to each following frame to estimate the motion of each feature and then update the tracking box according to the average estimated motion of all features. We used the OpenCV implementations of both algorithms with parameters listed in Table \ref{tab:params}\cite{opencv}.  Note that parameters were chosen qualitatively; for example, we chose the neighborhood size such that some visible cloud texture was encompassed within the neighborhood, though we expect the results to be robust for neighborhood sizes between $10$ and $20$ pixels for our studies.

In order to avoid tracking errors caused by corrupted image pixels, we implement a simple threshold on the data quality flag (DQF) field included in the ABI data files.  If the percentage of corrupted pixels in a frame surpassed 2\%, the frame was omitted from the optical flow computations.  This could be improved by restricting the threshold only to those pixels nearby the tracking box but we found our approach to be sufficient for our study as we encountered very few corrupted image frames. 

\begin{table}[!t]
	\caption{Shi-Tomasi and Lucas-Kanade parameter values chosen for selecting and tracking features in local cloud regions in GOES-17 imagery.  These parameters are inputs to the OpenCV v3 implementation of the algorithm called \texttt{goodFeaturesToTrack()} and \texttt{calcOpticalFlowPyrLK()}, respectively.   Refer to the OpenCV documentation for parameter descriptions.}
	\label{tab:params}
	\centering
	\begin{tabular}{|c|c|c|}
		\hline
		\textbf{Method} & \textbf{Parameter} & \textbf{Value}\\ \hline
		
		\multirow{6}{*}{Shi-Tomasi}	&maxCorners & 100\\ \cline{2-3}
								&qualityLevel & 0.2\\ \cline{2-3}
								&minDistance & 3\\ \cline{2-3}
								&blockSize & 3\\ \cline{2-3}
								&useHarrisDetector & False\\ \cline{2-3}
								&k & 0.04\\ \hline
		\multirow{4}{*}{Lucas-Kanade}	&winSize & $(15, 15)$\\ \cline{2-3}
								&maxLevel & 3\\ \cline{2-3}
								&criteria & 0.03 px or 10 iterations\\ \cline{2-3}
								&minEigThreshold & $10^{-4}$\\ \hline
	\end{tabular}
\end{table}

The LK motion estimation and tracking box updates continue (omitting corrupt frames as necessary) until a diurnal transition (sunrise or sunset), occurs.  Diurnal transition periods are determined via thresholds on the solar zenith angle, as described in Section \ref{sec:dtransitions}.
As the LK algorithm is sensitive to changes in pixel intensity between frames, it is unable to track features during the large intensity changes that occur during transitions between nighttime and daytime images.  Instead, we cease tracking and use a simple trajectory prediction method during these transitions.  Since the diurnal transition period is relatively short ($<$2 hr), we assume that the velocity of our tracking box is constant during this time and predict motion accordingly.  We estimate a constant-velocity trajectory by averaging the observed motion of the features over the six frames prior to a transition period and apply this linear trajectory to shift the tracking box during the transition.  Once the transition ends, we use the tracking box location to identify an updated ROI in the current frame.  We then again apply the ST algorithm within our ROI to identify trackable features followed by the LK algorithm to track those features in the following frames.
We find that this re-identification of good tracking features after a transition is especially helpful in the case of ship tracks as tracks change shape and stronger features become available than were originally chosen, intermittently recalibrating the LK algorithm.  

We continued this cycle of LK tracking and linear motion prediction for an initial 30 hr, which we found sufficient to capture the lifetimes of all the ship track intersections in our study, but ultimately disregarded those frames after which the ship track intersection was no longer recognizable. Track recognizability was determined qualitatively although could also be done quantitatively by comparing the track in frame $F_k$ to the original frame $F_0$ using the similarity metric we present in Section \ref{sec:results}.
Algorithm \ref{alg:of} summarizes our method for an ordered list of image frames $F$ observed at timestamps $t$ and user-selected ROI defined by location $tbox_{user}$, width $w$, and height $h$, in the first frame $F_0.$  Corrupt frames are identified with \texttt{DQ\_CHECK($F_k$, $T$)}, where $T$ is a threshold on the allowed percentage of corrupt image pixels; \texttt{ST($F_k$, $tbox_k$, $w$, $h$)} applies ST within the ROI defined by $tbox_k$, $w$, and $h$ in frame $F_k$ to identify feature locations $feats_k$; and \texttt{LK($F_k$, $F_{k+1}$, $feats_k$)} estimates the motion of the features at $feats_k$ between frames $F_k$ and $F_{k+1}$.

\algsetup{indent=2em}

\begin{algorithm}[H]
	\caption{Implementation of optical flow method.}
	\label{alg:of}
	\begin{algorithmic}
		\STATE
		\STATE \textbf{Begin:}
		\STATE $tbox \gets$ empty dictionary
		\STATE $feats \gets$ empty dictionary
		\STATE $tbox_0 \gets tbox_{user}$
		\STATE $feats_0 \gets$ \texttt{ST}$(F_0, tbox_0, w, h)$
		\STATE $i \gets 1$
		\WHILE {\textit{Time elapsed} $< 30 hr$ \AND $i < length(F)$} 
			\IF{\NOT \texttt{DQ\_check}$(F_i, 2)$}
				\STATE $i \gets i+1$
				\STATE continue
			\ENDIF
			\IF {\textit{Diurnal transition condition met}}
				\STATE $v \gets \frac{1}{5} \sum_{j=i-5}^{i-1}\frac{tbox_j-tbox_{j-1}}{t_j - t_{j-1}}$
				\WHILE {\textit{Diurnal transition condition met} \AND $i<length(F)$}
					\STATE $tbox_i \gets tbox_{i-1} + v\cdot(t_i - t_{i-1})$
					\STATE $feats_i \gets \text{None}$
					\STATE $i \gets i+1$
				\ENDWHILE
				\IF{$i \geq length(F)$}
					\STATE break
				\ENDIF
				\STATE $feats_{i-1} \gets$ \texttt{ST}$(F_{i-1}, tbox_{i-1}, w, h)$
			\ENDIF
			\STATE $feats_i \gets$ \texttt{LK}$(F_{i-1}, F_i, feats_{i-1})$
			\STATE $tbox_i \gets tbox_{i-1} + \frac{\sum(feats_i - feats_{i-1})}{length(feats_i)}$
			\STATE $i \gets i+1$
		\ENDWHILE
	\end{algorithmic}
\end{algorithm}

\begin{figure*}[!t]
	\centering
	\subfloat{\includegraphics[width=1.88in]{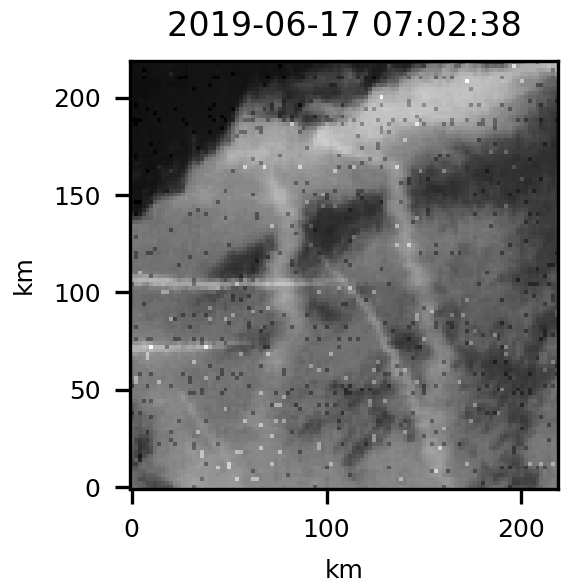}}
	\subfloat{\includegraphics[width=1.88in]{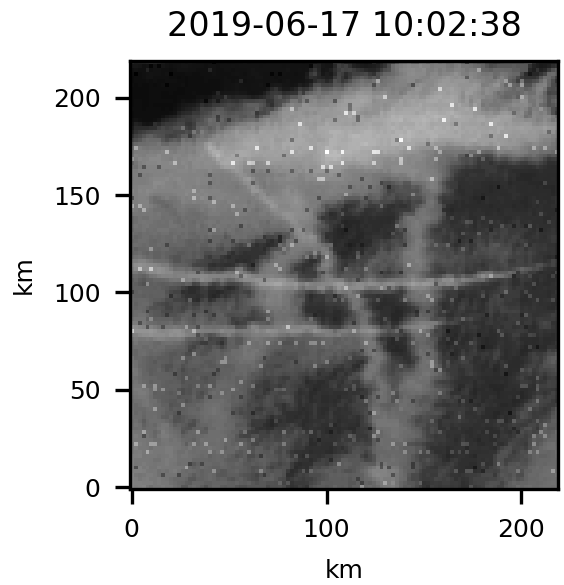}}
	\subfloat{\includegraphics[width=1.88in]{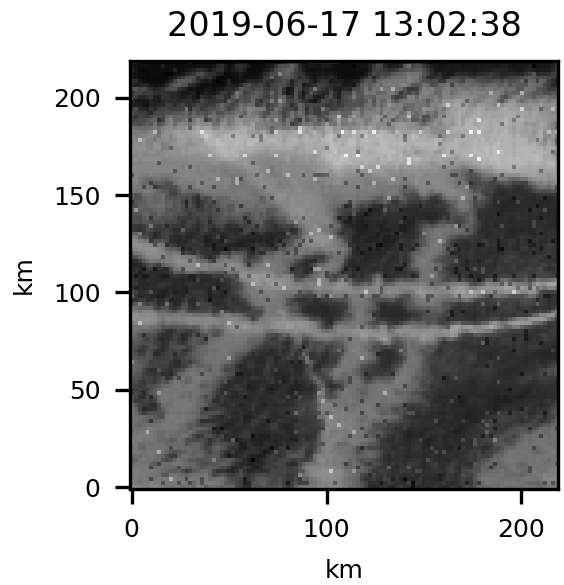}}\\
	\subfloat{\includegraphics[width=1.88in]{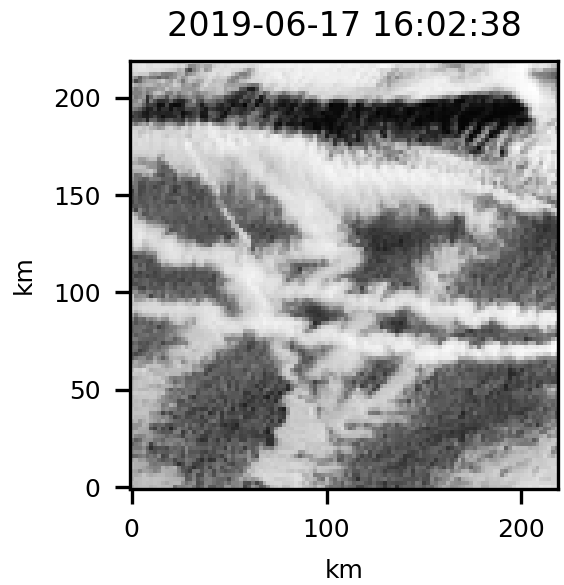}}
	\subfloat{\includegraphics[width=1.88in]{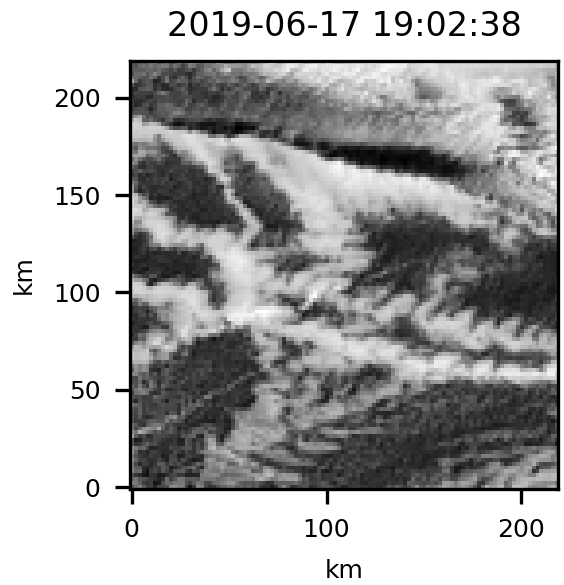}}
	\subfloat{\includegraphics[width=1.88in]{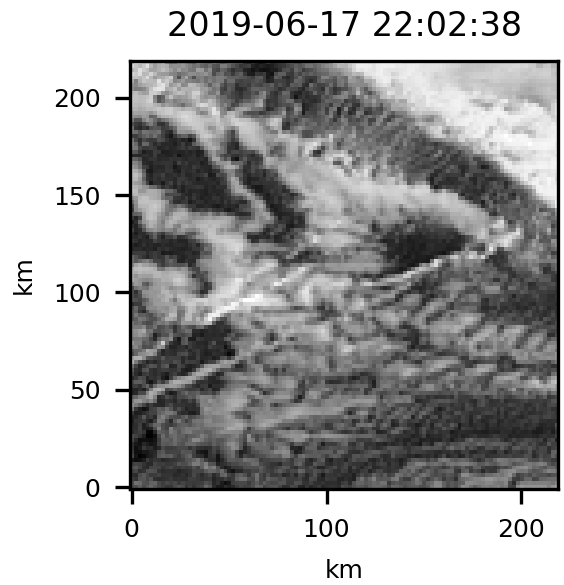}}\\
	\subfloat{\includegraphics[width=1.88in]{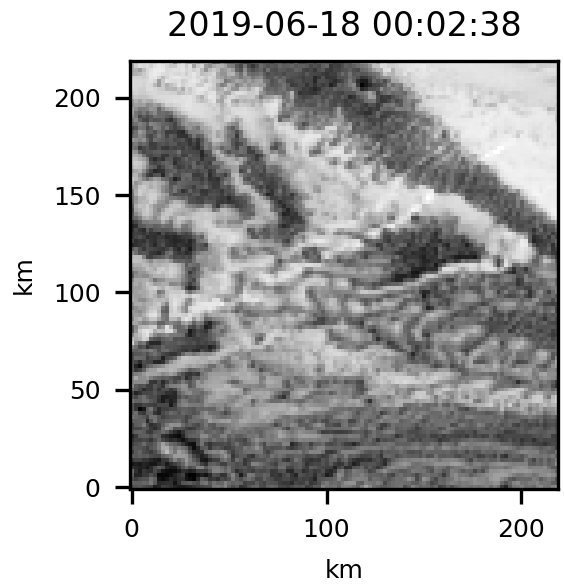}}
	\subfloat{\includegraphics[width=1.88in]{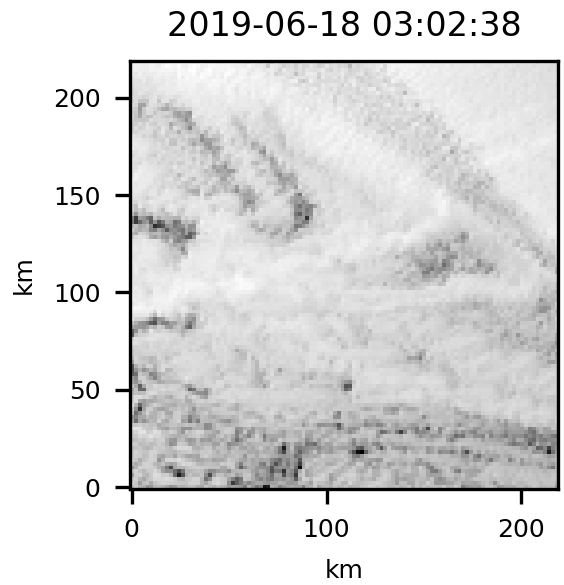}}
	\subfloat{\includegraphics[width=1.88in]{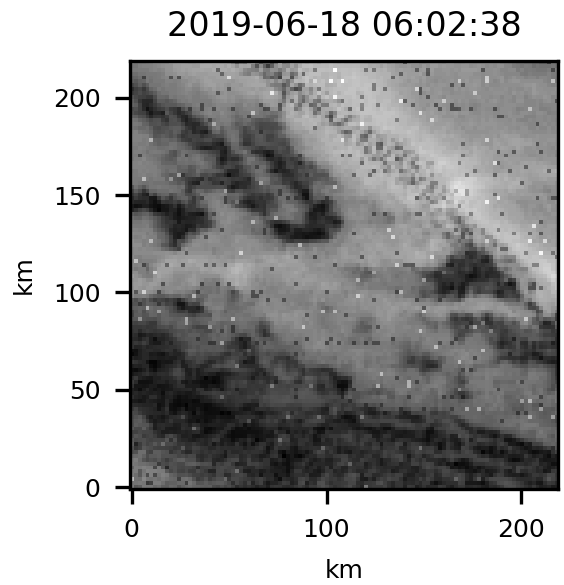}}
	
	\caption{These figures show the 110 px by 110 px image region centered on the tracking box of the optical flow method at each frame in 3-hr time intervals beginning on June 17, 2019, at 07:02 UTC. The intersection of interest is at the center of each image. The tracking box is removed from these images for better visualization.  The remnants of the ship tracks are still visible up to 20 hr after they first appeared.  The center locations of these images in order from left to right then top to bottom is as follows: 36°11'14.1"N 134°15'14.8"W; 35°37'10.0"N 134°25'36.5"W; 34°57'23.3"N 134°38'39.4"W; 34°16'38.2"N 135°00'31.1"W; 33°36'15.9"N 135°31'41.3"W; 32°53'41.3"N 136°06'49.2"W; 32°24'26.7"N 136°35'03.9"W; 31°44'39.5"N 137°25'12.0"W; 31°12'18.7"N 138°21'42.1"W.}
	\label{fig:lk_june_small}
\end{figure*}

Although best assessed in video format, we show an example application of this method with a case study from June 17-18, 2019.  Fig. \ref{fig:lk_june_big} shows the tracking box following an intersection of two distinct ship tracks over 18 hours, throughout which the tracks clearly persist.  Fig. \ref{fig:lk_june_small} shows the isolated ROI of the same tracking result in 3-hr increments over a 23-hr period.  We clearly observe ship tracks from the beginning of tracking at 07:02 UTC until at least 03:02 UTC the following day, quantifying the duration of persistence as 20 hr. Using this approach, we can observe the persistence and dispersion of ship tracks in a low cloud layer by tracking that region over many hours in increments of 5 min, the temporal resolution of the GOES-17 ABI CONUS scan. Note that this tracking algorithm would also work in increments of 10 or 15 minutes but at the cost of tracking accuracy that is not worth the negligible efficiency gained. It is recommended to use time intervals as close as possible. 

\subsection{Details on diurnal transition timing using solar zenith angle}\label{sec:dtransitions}

We determine the starting and ending frames of a sunrise or sunset transition period via comparisons of solar zenith angle calculated along the right and left edges of the tracking box in each frame.  These angles were calculated using the National Renewable Energy Laboratory’s Solar Position Algorithm \cite{nrel2008}, which calculates the sun's apparent altitude with a precision of about 0.0003 degrees given the date, time, and location.  For a given frame, the minimum and maximum angles are used for both the right and left edges of the tracking box to determine the beginning and end of a diurnal transition.  Thresholds for the start and end of each transition were chosen empirically and conservatively to ensure the transition periods are estimated accurately.

Specifically, let $\alpha_r$ and $\alpha_l$ be vectors of pixel solar zenith angles for the right and left half of the tracking box perimeter, respectively, and let $c, d \in \mathbb{R}$ such that $0< c < d$ be the thresholds for transition periods. Then a diurnal transition is occurring if the following condition is true:

\begin{equation*}
(\hbox{min}(\alpha_r) < d \cap \hbox{max}(\alpha_l) > c ) \cup (\hbox{max}(\alpha_r) > c \cap \hbox{min}(\alpha_l) < d ).
\end{equation*}

\begin{figure*}[!h]
	\centering
	\includegraphics[trim = 0cm 0cm 0cm 0cm, clip, width=5in]{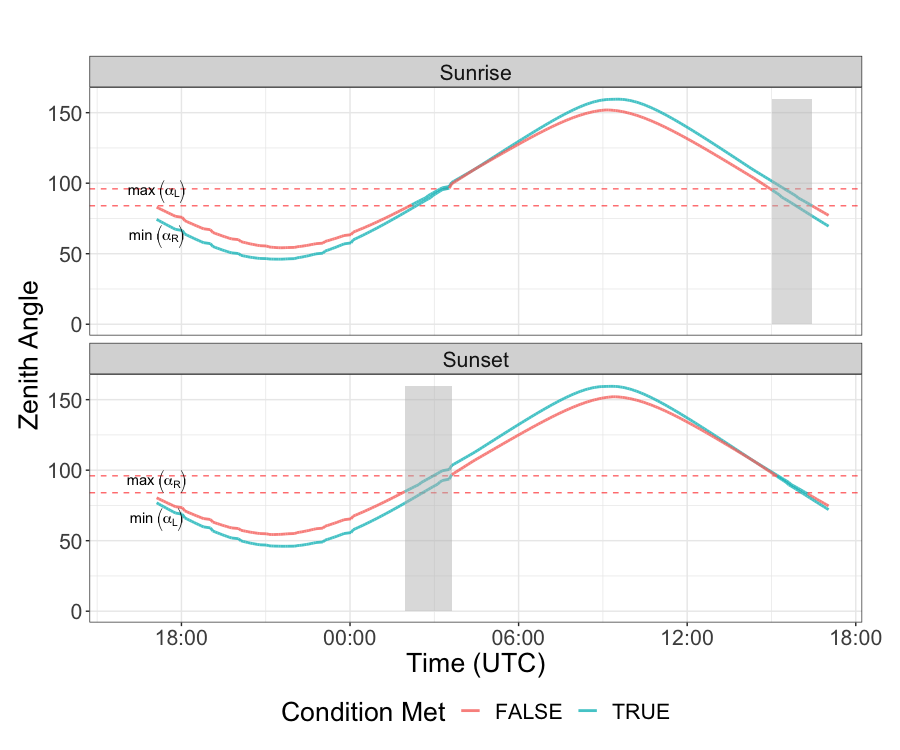}
	\caption{This figure demonstrates the diurnal cycle of the minimum and maximum solar zenith angles for the right and left side of the tracking box over a 25 hour period starting on February $20^{th}$ at 17:00 UTC and ending on February $21^{th}$ at 18:00 UTC. The top figure shows the behavior of $\hbox{max}(\alpha_L)$ $\hbox{min}(\alpha_R)$, which indicate the start and stop of a sunrise transition while the bottom figure shows the behavior of $\hbox{max}(\alpha_R)$ $\hbox{min}(\alpha_L)$, which indicate the start and stop of a sunset transition. The vertical grey boxes indicate the time period where the transition conditions are met and thus the time periods for a sunrise or sunset event}
	\label{fig:transitions}
\end{figure*}

The first set in the union describes a sunrise and the second describes a sunset.
A demonstration of this condition is displayed in Fig. \ref{fig:transitions}, which shows the diurnal pattern of rising and falling slopes of $\hbox{max}(\alpha_l)$, $\hbox{min}(\alpha_r)$, $\hbox{max}(\alpha_r)$, and $\hbox{min}(\alpha_l)$ for the February case study. 
The sunrise boundary begins at the frame in which $\hbox{min}(\alpha_r) < d$ on a falling slope of $\hbox{min}(\alpha_r)$ values and ends at the frame where $\hbox{max}(\alpha_l) \leq c.$
The sunset boundary begins at the frame in which $\hbox{max}(\alpha_r) > c$ on a rising slope of $\hbox{max}(\alpha_r)$ values and ends at the frame where $\hbox{min}(\alpha_l) \geq d$.

\begin{figure*}[!h]
\centering
\subfloat[Sunrise]{\includegraphics[width=2.75in]{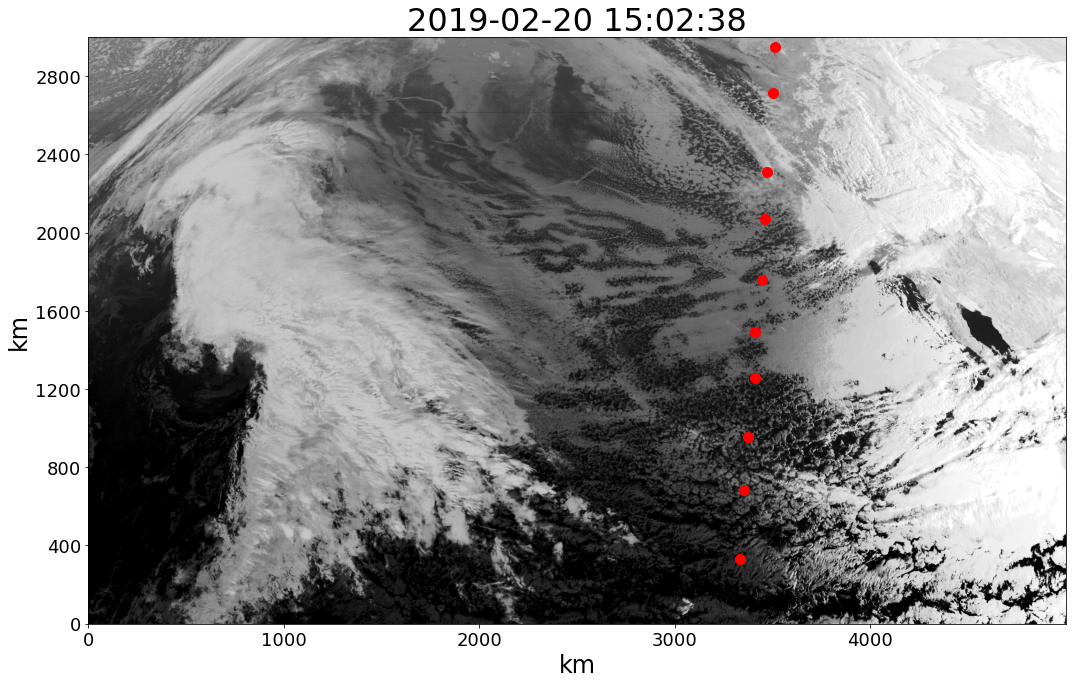}}
\subfloat[Sunset]{\includegraphics[width=2.75in]{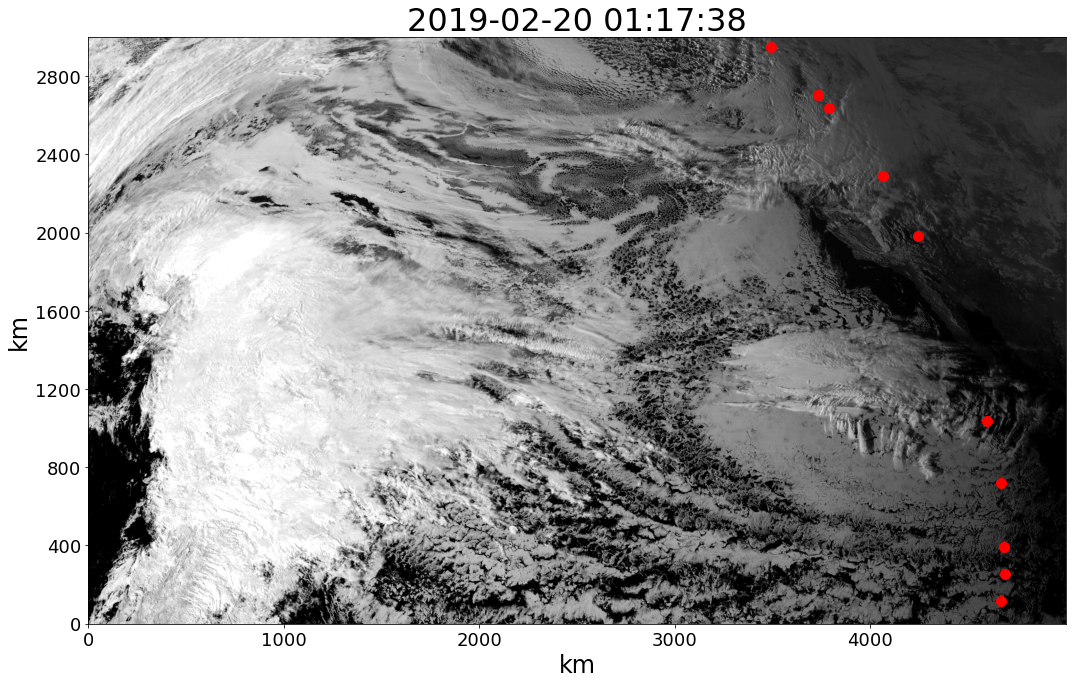}}

\caption{Examples of selected observations (red points) for a sunrise (a) and sunset (b) transition on February $20^{th}$.  The center location of these images is 29°57'19.4"N 136°59'21.6"W.}
\label{fig:transition_points}
\end{figure*}

To determine empirical values for $c$ and $d$, we sampled a total of eight sunrise and nine sunset images from multiple dates. Using ten points manually selected from each image where sunrise or sunset significantly impacted cloud radiance, we derived a total of 160 solar zenith angles. Figs. \ref{fig:transition_points} (a) and (b) show an example of selected points along the start of a sunrise and sunset transition, respectively.

We initially chose sunrise and sunset thresholds as the 95th percentile of the angles sampled for each, then manually tuned the values during application to improve the estimate.  Ultimately we chose $d=97^\circ$ and $c=83^\circ$; we found these values to be robust for our study.

\section{Comparison with HYSPLIT}
\label{sec:hysplit}
\noindent We employ NOAA’s HYSPLIT model \cite{hysplit} as a baseline method for comparison used to follow ship tracks. Given HYSPLIT's popularity for trajectory analysis, it is, in theory, able to capture the atmospheric movement of air parcels that make up a specific ship track, showing where a given feature - in this case, an intersection of two ship tracks - moves in subsequent time periods. We run simulations using analysis data from the Global Data Assimilation System (GDAS) provided by the National Weather Service's National Centers for Environmental Prediction. GDAS data is often used when gridded observational data is required, and the Air Resources Laboratory processes this data into HYSPLIT-useable formats. 
Prior to June 12, 2019, GDAS data had a spatial resolution of 0.5$^\circ$ grid cells ($\approx$ 50 km), and subsequent dates have a resolution of 0.25$^\circ$ grid cells ($\approx$ 25 km). Our simulation study relies on data both before and after that date (both resolutions). For all trajectory analyses, the temporal resolution is hourly.
 
\begin{figure*}[!t]
\centering
\includegraphics[width=2.0in]{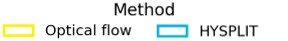}
\\
\subfloat[Initialization: 0 Hours]{\includegraphics[width=2.75in]{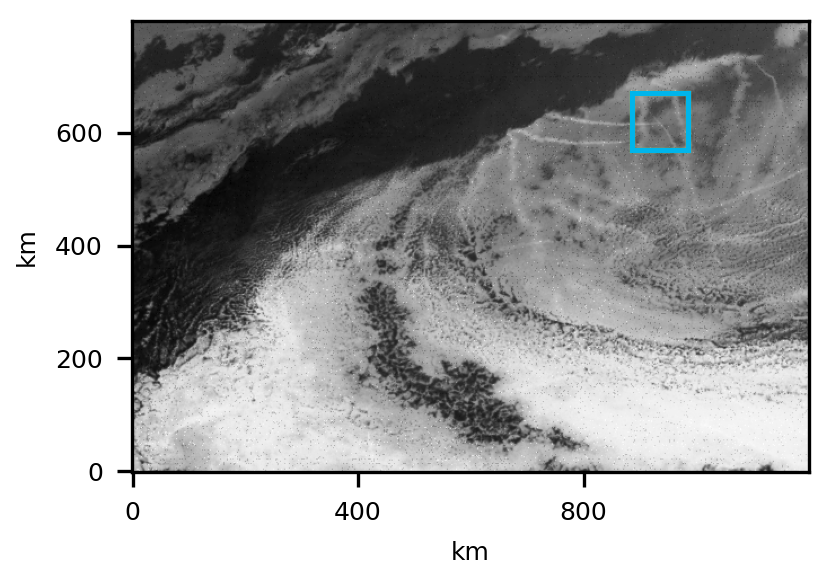}}
\subfloat[6 Hours]{\includegraphics[width=2.75in]{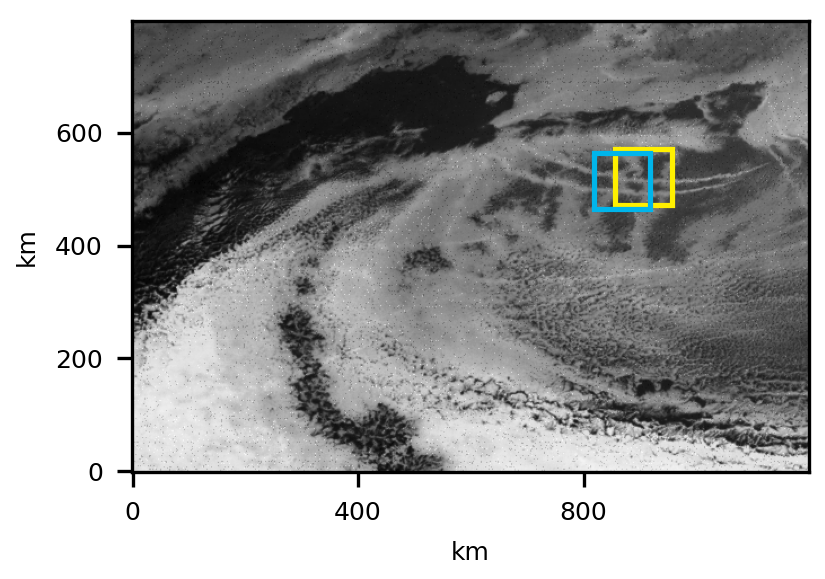}}\
\subfloat[9 Hours]{\includegraphics[width=2.75in]{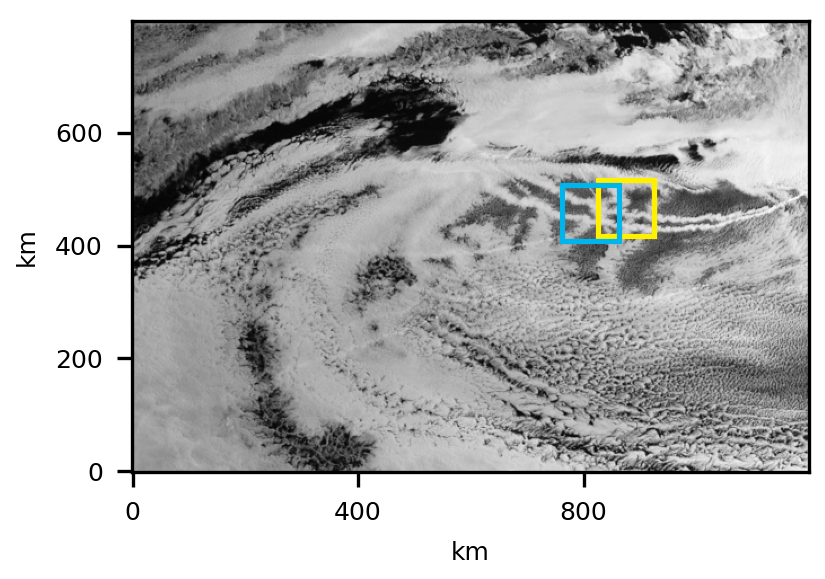}}
\subfloat[12 Hours]{\includegraphics[width=2.75in]{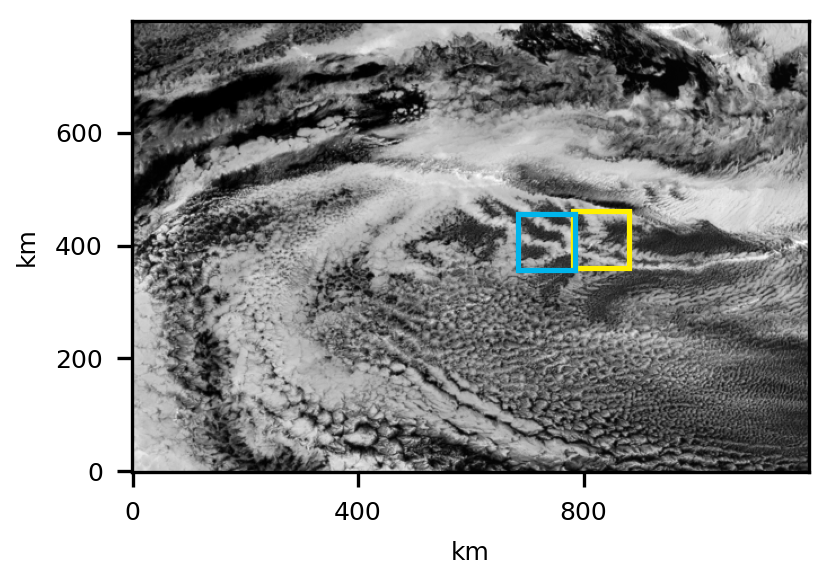}}

\caption{These figures show the forward trajectory of air parcel movement, starting with the head of a track on June 17, 2019, at 07:02 UTC (a) and stepping forward in time, with snapshots shown 6 (b), 9 (c), and 12 (d) hr later. HYSPLIT is able to track the movement of this cloud feature well, and the track is still clearly visible at 12 hr.  The center location of these images is 33°27'02.0"N 138°06'11.9"W.}
\label{fig:of-hysplit_june}


\end{figure*}

We use HYSPLIT to project the movements of pre-formed ship tracks forward in time. We initialize the trajectory analysis for a specified intersection of two ship tracks.
We then use HYSPLIT to compute the forward trajectory for an air parcel using these positions as the initial location and let the simulation run 24 hr forward in time. 
For a specified intersection of ship tracks, HYSPLIT was initialized at the coinciding timestamp, location and initialization height. 
Initialization height was determined using the boundary layer height as collected from ERA5 reanalysis data for that location \cite{hersbach2020era5}.  With the trajectory points, we then overlaid the HYSPLIT predictions onto the satellite imagery for these forward time points and visually assessed a) how well the air parcel projections estimated the observed ship track movement and b) whether there was still a clear remnant of the ship track intersection at each forward time step.  Fig. \ref{fig:of-hysplit_june} shows what this looks like for an initialization point from June 17, 2019 compared with using the optical flow approach described in Section \ref{sec:imageflow}.
Overall, the predicted HYSPLIT trajectories estimated the ship track movement reasonably well within the first 8 to 12 hr, but the errors inherent in the model are likely too large to capture the precision needed for following a such small cloud structures. There is likely uncertainty in the initialization height, as we do not know for sure at which elevation the aerosol-brightened clouds exist. While HYSPLIT captures the overall atmospheric movement well and can predict the direction and scale of ship track movement, it fails to follow a specific ship track intersection with sufficient precision. 
%


In general, we found HYSPLIT to be useful in predicting the trajectories of visible ship tracks up to approximately 8--12 hr depending on the case. Given the location errors introduced when tracking small features, HYSPLIT provides only a rough estimate of track persistence, and we also lose confidence in our ability to identify the original track as it evolves. For ship tracks that are less distinguishable from the surrounding clouds, it becomes difficult to confirm that HYSPLIT projections align with the same portion of the track where we initialized the forward trajectory. This is likely due to the fact that 1) HYSPLIT is initialized at a single location and then runs forward with no intermediate checks to see if it is still tracking the same (initial) feature and 2) HYSPLIT only projects at 25 km spatial and 1-hr increments (compared to the 0.5--2 km and 5--15 minute resolutions the GOES-17 imager collects) and might miss nonlinear movement within the hour.


\section{Results}
\label{sec:results}
To demonstrate the performance of our optical flow algorithm for tracking ship track segments in low-lying clouds, we compare our method to using NOAA's HYSPLIT model \cite{hysplit} to track the 24 intersection points as described in Section \ref{sec:data} with details provided in Table \ref{tab:intersections}.  Using identifiable intersections allows us to qualitatively assess the precision of the feature tracking. Since HYSPLIT predicts trajectories in hourly increments, we chose initialization points as close to the top of the hour as possible. We initialized our optical flow approach at these times as well for consistency and to appropriately compare results.

To quantitatively assess the accuracy, we compute the structural similarity index measure (SSIM) \cite{wang2004} to measure the similarity between tracking boxes (ROIs) identified in subsequent image frames.  SSIM measures perceived changes in structural information.  We expect as persistence prolongs, the similarity between tracking boxes centered around a ship track feature will naturally degrade no matter what method is used for tracking.  Higher SSIM values indicate better performance.

\begin{figure}[!t]
\centering
\subfloat[]{\includegraphics[width=0.325\linewidth]{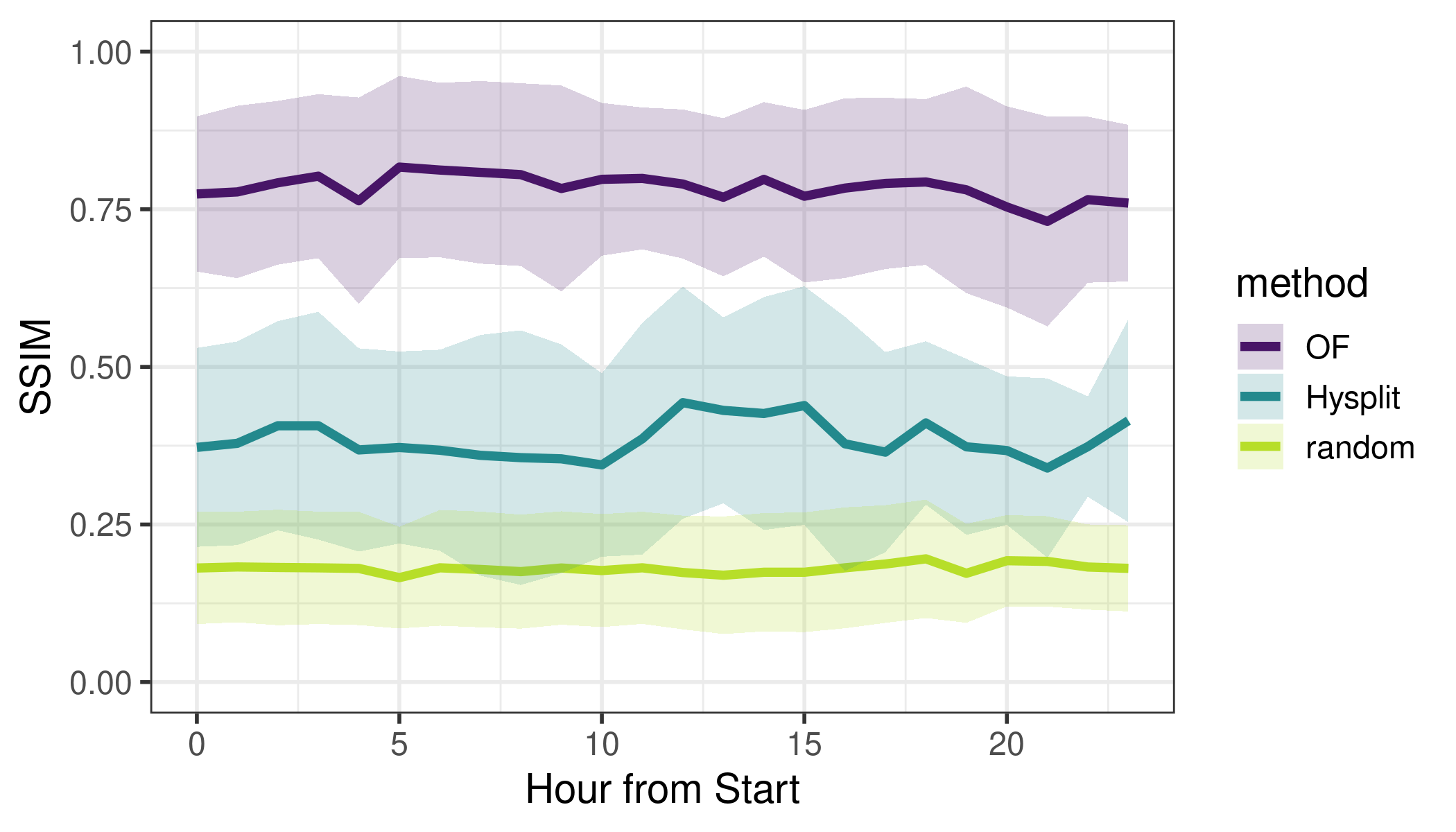}}
\subfloat[]{\includegraphics[width=0.325\linewidth]{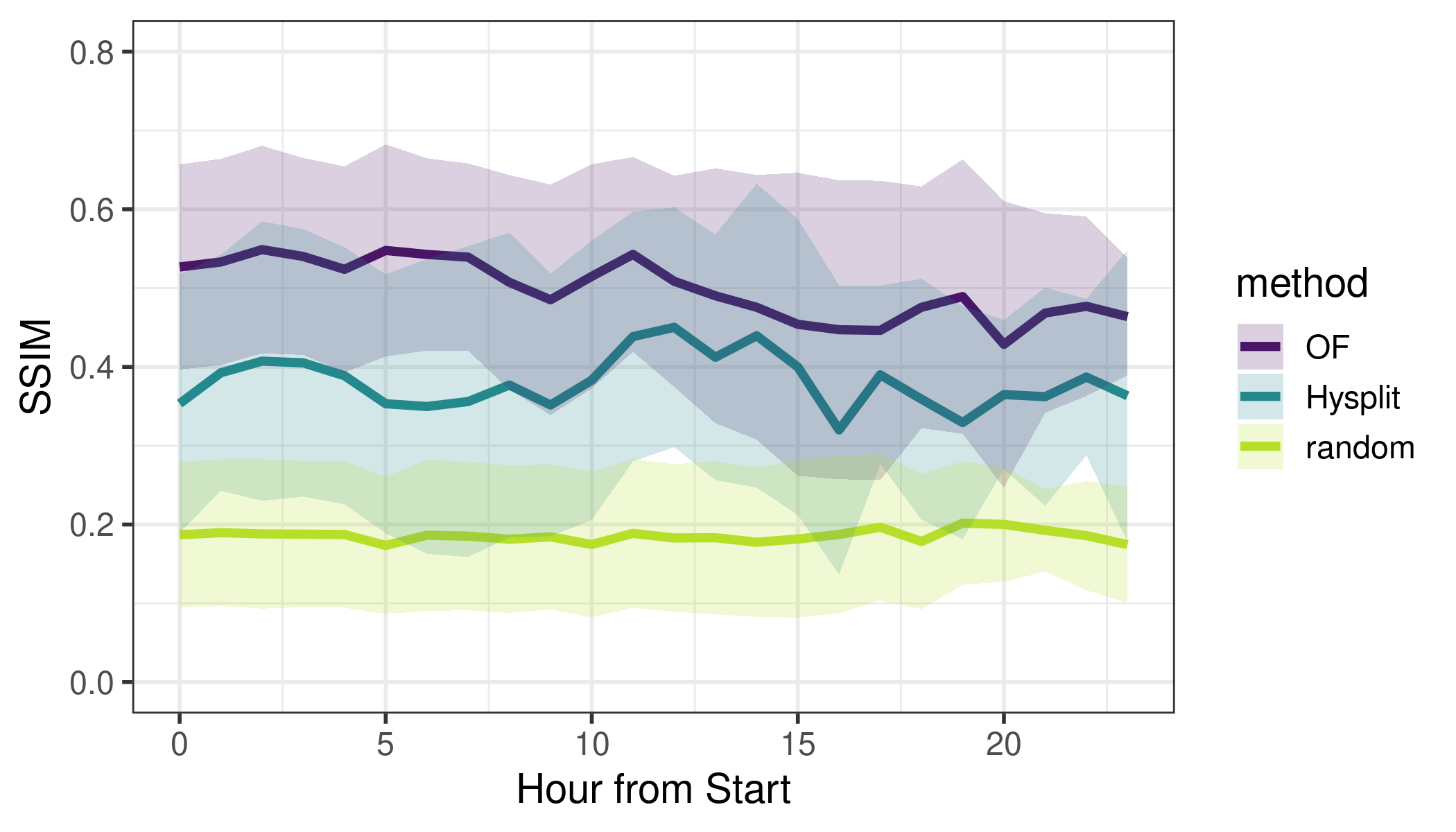}}
\subfloat[]{\includegraphics[width=0.325\linewidth]{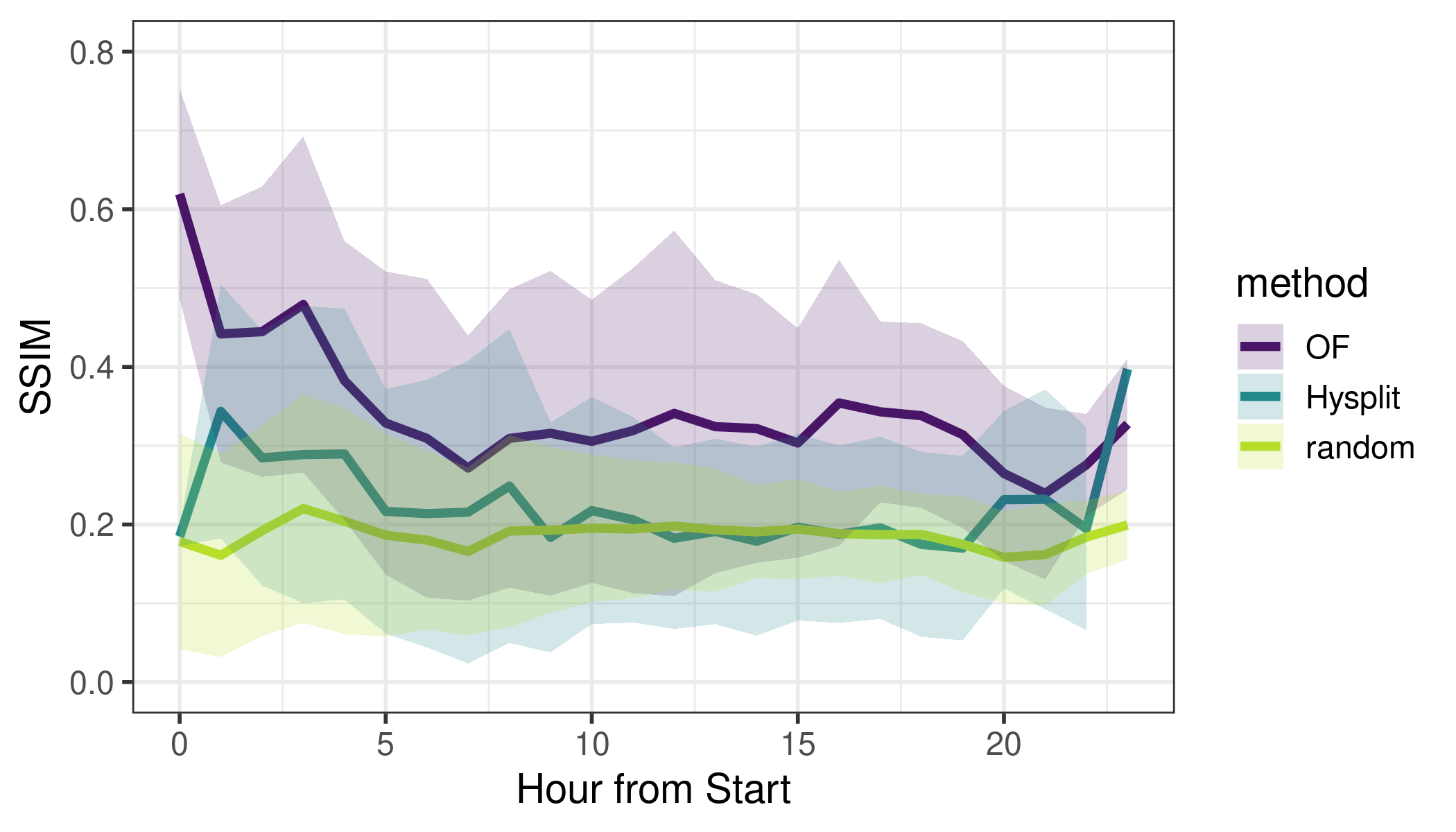}}

\caption{(a) Frame by frame comparison of HYSPLIT performance to Optical Flow performance at 5-min and (b) hourly intervals. HYSPLIT performance is only computed at hourly intervals. (c) Reference frame comparison of HYSPLIT performance to Optical Flow at hourly intervals.
Solid lines represent a smoothed mean SSIM across all samples with transparent shading representing one standard deviation from the mean.}
\label{fig:results}
\end{figure}

Since HYSPLIT only produces projections at hourly time stamps, we are only able to measure HYSPLIT's accuracy at a minimum of hourly intervals. To appropriately compare the performance of our approach, we compared images 1 hour apart. Note that we can compute SSIM at time intervals as short as 5 minutes since our approach relies on GOES-R data but this is not possible with HYSPLIT. Due to the shorter time intervals, the optical flow method has a biased advantage when comparing performance with correlation-based similarity metrics such as SSIM. We show optical flow results considering both 5-min (Fig. \ref{fig:results} (a)) and hourly intervals (Fig. \ref{fig:results} (b)) to demonstrate this bias.

Lastly, to show strong evidence we are tracking the same initial intersection in frame one, we compute SSIM between each frame and a ``true" reference frame. Ideally, the reference frame is the first tracking frame. However, due to the shift in image brightness and texture after a diurnal transition, we select new features within the tracking box after each transition. Thus, the reference frame is the first tracking frame until a transition occurs, in which case the reference frame becomes the first frame after a transition. Figure \ref{fig:results} (c) compares each frame to the reference frame.

Overall, our approach, denoted \texttt{OF} in Fig. \ref{fig:results}, clearly outperforms the use of HYSPLIT to predict ship track trajectories. We also compare our results to randomly selected same size boxes in subsequent frames, denoted as \texttt{random} in Fig. \ref{fig:results}, as another comparison baseline and sanity check.
As expected, the performance of our optical flow approach slightly degrades over time as ship tracks become no longer distinguishable from the background clouds. This performance degradation is seen by the slight but noticeable decrease in SSIM after the 20 hr mark.

\section{Discussion}
\label{sec:disc}
We have presented a novel method of systematically following ship track behavior observed in imagery from the GOES-17 geostationary weather satellite. Our approach can follow a ship track with high accuracy well past 12 hr and throughout day-night transitions, enabling more precise characterization of ship track persistence. We demonstrated ship track tracking capabilities over 24 key case studies and shown it outperforms the standard HYSPLIT trajectory analysis tool. The high temporal resolution of the GOES-R ABI allows us to assume minimal changes between image frames even though the form of the ship track evolves over time.  A track that is 3 or 6 hours old will have a different shape than when the track initially forms making it previously impossible to track ship tracks using imagery alone prior to the availability of the new GOES-R satellite series. This dynamic track evolution prevents the use of data collected by most other satellite instruments such as MODIS, for tracking.

Although we demonstrated the efficacy of the optical flow approach, it is important to discuss some of its limitations as well as potential areas of future study.  Most feature tracking errors stem from the method's assumption that the selected features appear roughly similar from one frame to the next, making it sensitive to artifacts such as data corruption, abrupt changes in pixel intensity, and non-affine feature warping.

While we found our method of predicting tracking box motion during diurnal transitions to be successful in our study, one can imagine realistic cases where the assumption of constant velocity is false; for example, a cold front may cause an abrupt shift in wind patterns.  To avoid the prediction of anticipated changes due to diurnal transitions, it would be advantageous to remove the intensity variations all together in a preprocessing step. This would allow us to leverage the LK method for the full 24 hr period. The challenges that pixel intensity variations can bring are often nuanced, application specific and nontrivial to overcome.  Techniques for correcting intensity inhomogeneities in magnetic resonance imaging (MRI) brain scans could be leveraged for this step. Intensity non-uniformity (INU) correction methods rely only on image features to remove spatial inhomogeneities and have been shown to be effective (e.g. \cite{ganzetti2016} and references therein).


In addition to changes in pixel intensity, the optical flow method is also sensitive to non-affine changes in the shapes of features.  Common cloud motions can introduce warping, but the warping between two consecutive GOES-17 CONUS frames, measured 5 min apart,  is minimal and does not affect tracking success in our selected test cases.  A larger temporal gap between frames may lead to greater observed changes in the shape of a ship track  and thus introduce notable tracking errors.  Large temporal differences between frames can occur when data is absent from the database or the frames are rejected because they contain corrupt data.  In our experience, a temporal gap of up to one hr between two frames is generally reasonable for successful feature tracking.

Other natural phenomena that can cause the Lucas-Kanade technique to fail include interference from high-altitude clouds passing over the region of interest, dispersion of the boundary cloud layer, and disappearance of texture from the cloud layer.  
In the first case, the features may be obscured or confused with similar features in the higher-altitude clouds; in the latter two cases, the features disappear entirely and cannot be tracked further.
Many of the obstacles described here could be circumvented by integrating known physical and/or meteorological factors that contribute to cloud feature movement. 
There is also the potential to integrate the HYSPLIT tracking approach described in Section \ref{sec:hysplit} over short periods of time ($<$6 hr) when we expect the optical flow approach to fall short.

\paragraph{Acknowledgments}
We would like to acknowledge Rob Wood and the atmospheric sciences department at University of Washington, conversations with whom helped propel this work forward. We would also like to thank Katherine Simonson at Sandia National Lab for her contributions to this work and Lori Dotson for contributing to the editing process.

\paragraph{Funding Statement}
This paper describes objective technical results and analysis. Any subjective views or opinions that might be expressed in the paper do not necessarily represent the views of the U.S. Department of Energy or the United States Government. This work was supported by the Laboratory Directed Research and Development program at Sandia National Laboratories, a multi-mission laboratory managed and operated by National Technology and Engineering Solutions of Sandia, LLC, a wholly owned subsidiary of Honeywell International, Inc., for the U.S. Department of Energy's National Nuclear Security Administration under contract DE-NA0003525. SAND2022-9069 J


\bibliographystyle{unsrtnat}
\bibliography{biblio}  

\end{document}